\documentclass[amsmath,superscriptaddress,prl,twocolumn] {revtex4}
\usepackage{bm}
\usepackage{enumerate}
\usepackage{graphicx}
\usepackage[dvips]{epsfig}
\usepackage{epsf}
\usepackage{xcolor}

\makeatletter
\newcommand{\Rmnum}[1]{\expandafter\@slowromancap\romannumeral #1@}
\makeatletter

\begin{document}

\title{Surface weak ferromagnetism}

\author{Vladimir A. Zyuzin}
\affiliation{L.D. Landau Institute for Theoretical Physics, 142432, Chernogolovka, Russia}

\begin{abstract}
This paper proposes a model of a N\'{e}el-ordered antiferromagnet exhibiting a weak ferromagnetic moment restricted to the crystal surface, generated by spin-orbit coupling induced by broken surface symmetries. Depending on crystal termination, the surface magnetic moment can align with identical or opposing signs on the top and bottom surfaces, resulting in a zero Faraday effect paired with a finite Kerr effect upon reflection for the latter configuration.
\end{abstract}
\maketitle

According to conventional classification of collinear antiferromagnets \cite{Dzyaloshinskii1958,Turov1965}, all N\'{e}el-ordered magnets can be divided into three categories: genuine antiferromagnets, weak ferromagnets, and ferrimagnets (a comment is in order \cite{commentA,commentB}). Genuine antiferromagnets have zero net magnetization for any direction of the N\'{e}el order, enforced by symmetries that relate the magnetic sublattices. Weak ferromagnets \cite{BorovikRomanov1957,Dzyaloshinskii1958,Moriya1960b} are collinear antiferromagnets in which the magnetic unit cells are likewise related by symmetry operations. However, in weak ferromagnets, the crystal symmetry and certain orientations of the N'{e}el order can permit a finite net magnetization despite the underlying compensated magnetic order.
The magnetization might get developed in the canting of the N\'{e}el order, orbital magnetization of fermions that interact with the N\'{e}el order, or orbital magnetization of spin-waves (magnons) that are fluctuations about the N\'{e}el order. Importantly, among the other restrictions, the symmetry class of weak ferromagnets must not have a combination of the time-reversal together with translation by the half of the unit cell or inversion in the list of symmetry operations that connect the magnetic sublattices of the N\'{e}el order. The remaining symmetry operations that connect the magnetic sublattices are typically combinations of time-reversal together with rotation and/or mirror operations. 
It is convenient to describe a compensated N\'{e}el order's ability to result in a finite magnetization in terms of a Dzyaloshinskii invariant\cite{Dzyaloshinskii1958,Turov1965,VonsovskiiTurov1986,Turov1990} (see \cite{GolubinskiiZyuzin2026,commentRutile1} for a recent review). Weak ferromagnets for all the symmetry classes are classified in the table 3 in \cite{Turov1965}. 
Examples of weak ferromagnets include $\alpha-$Fe$_2$O$_3$, LuFeO$_3$, YFeO$_3$, MnTe, NiF$_2$ and many more \cite{BorovikRomanov1957,BorovikRomanov1960,Dzyaloshinskii1958,Turov1965,VonsovskiiTurov1986,BorovikRomanov1973,Turov1990}. It can be shown that Dzyaloshinskii's invariant of weak ferromagnets have magnetization perpendicular to the N\'{e}el vector. However, a combined effect of different invariants might make the overall magnetization to have a component also along the overall N\'{e}el \cite{commentRutile1} vector.

Finite magnetization in weak ferromagnets occurs as a result of spin-orbit coupling that is allowed by the symmetry of the crystal. In case of orbital magnetization, in addition to the spin-orbit coupling, a particular anisotropic structure of fermion tunneling in the fermionic case or exchange interaction in case of magnons, that are allowed by the symmetry of the crystal, is typically also required for the magnetization to occur. This anisotropic structure was shown to result in antiferromagnetic spin-splitting \cite{Noda2016,Okugawa2018,Naka2019,HayamiYanagiKusunose2019} of either fermions or magnons. 
We note that spin-splitting was experimentally, indirectly observed in CoF\({}_{2}\) through the quadratic-in-magnetic-field Faraday effect \cite{KharchenkoBibikEremenko1985,EremenkoKharchenko1987}. This phenomenon was ascribed strictly to the underlying structural symmetry of the CoF\({}_{2}\) antiferromagnet \cite{VonsovskiiTurov1986,Turov1990}. Later the d-wave Hall effect consistent with the observation of \cite{KharchenkoBibikEremenko1985} was theoretically derived from microscopic model of d-wave antiferromagnetic spin-splitting \cite{VorobevZyuzin2024}.

The anomalous Hall effect due to the N\'{e}el vector in collinear antiferromagnets was theoretically discussed in \cite{TurovShavrov}, and has been experimentally discovered in $\alpha$-Fe$_2$O$_3$ \cite{Vlasov}. Recent experiments have successfully confirmed that the anomalous Hall effect in antiferromagnet MnTe is due to the N\'{e}el vector and its unique weak ferromagnetic crystal structure \cite{MnTeExp2023,MnTeExp2024,CommentMnTe}.
This behavior is supported by microscopic calculations of the Hall effect in collinear antiferromagnets using the Berry curvature framework \cite{Solovyev1997,Naka2020,AHE_AFM,Zyuzin2024,Zyuzin2025a,Solovyev2025,GolubinskiiZyuzin2026,commentRutile2}. These modern calculations firmly validate foundational physics theories established decades ago \cite{TurovShavrov,Vlasov}.
The Faraday effect in a collinear antiferromagnet, which is closely related to the anomalous Hall effect and driven by the N\'{e}el vector, was experimentally observed in YFeO\({}_{3}\) \cite{Krichevtsov1981}. This observation confirms that a collinear, compensated N\'{e}el order can give rise to orbital magnetization without requiring any canting of the magnetic order.
Finally, the spin-splitting in antiferromagnets have been shown to result in generation of spin currents \cite{Naka2019,Exp2026,DevyatovExp2026} and spin anomalous Hall effect \cite{Zyuzin2025a}. These effects contribute to the reseach field of spintronics \cite{Dyakonov}.

Ferrimagnets are N\'{e}el-ordered magnets characterized by a lack of symmetry connecting their magnetic sublattices. Consequently, these systems can exhibit a finite spontaneous magnetization, provided a Dzyaloshinskii invariant is permitted by the crystal symmetry operations. Recent studies \cite{Zyuzin2026,GolubinskiiZyuzin2026} demonstrate that ferrimagnets can be classified into two distinct groups based on the nature of this Dzyaloshinskii invariant. The first group comprises longitudinal ferrimagnets, which, in contrast to weak ferromagnets, possess a magnetization parallel to the N\'{e}el vector. The second group consists of transverse ferrimagnets, where the magnetization is perpendicular to the N\'{e}el vector, mimicking the behavior of weak ferromagnets. Depending on the spatial orientation of the N\'{e}el vector, certain ferrimagnets may exhibit characteristics of both types simultaneously. Furthermore, the resulting spin splitting and anomalous Hall effect have been theoretically investigated within simple models spanning both classes of ferrimagnets \cite{Zyuzin2024,Zyuzin2025a,Zyuzin2026,GolubinskiiZyuzin2026}.

In this paper, we propose and theoretically investigate a minimal model of a hitherto unexplored form of weak ferromagnetism that occurs exclusively on the surfaces of a N\'{e}el-ordered antiferromagnet. In the bulk system, the Dzyaloshinskii invariant required for a net magnetic moment vanishes due to a specific combination of symmetry operations. However, the spatial termination of the crystal breaks these symmetries, stabilizing a net magnetic moment solely at the surface boundaries. While uncompensated ferromagnetic magnetization can occur at the surfaces or dislocations of conventional layered antiferromagnets, such as CrSb, MnTe, or families of Eu-based compounds \cite{EuAFMPudalov} where individual layers are ferromagnetically ordered, the mechanism of surface weak ferromagnetism proposed here arises from a fundamentally different physical origin.

\begin{figure}[t] 
\includegraphics[width=0.85 \columnwidth ]{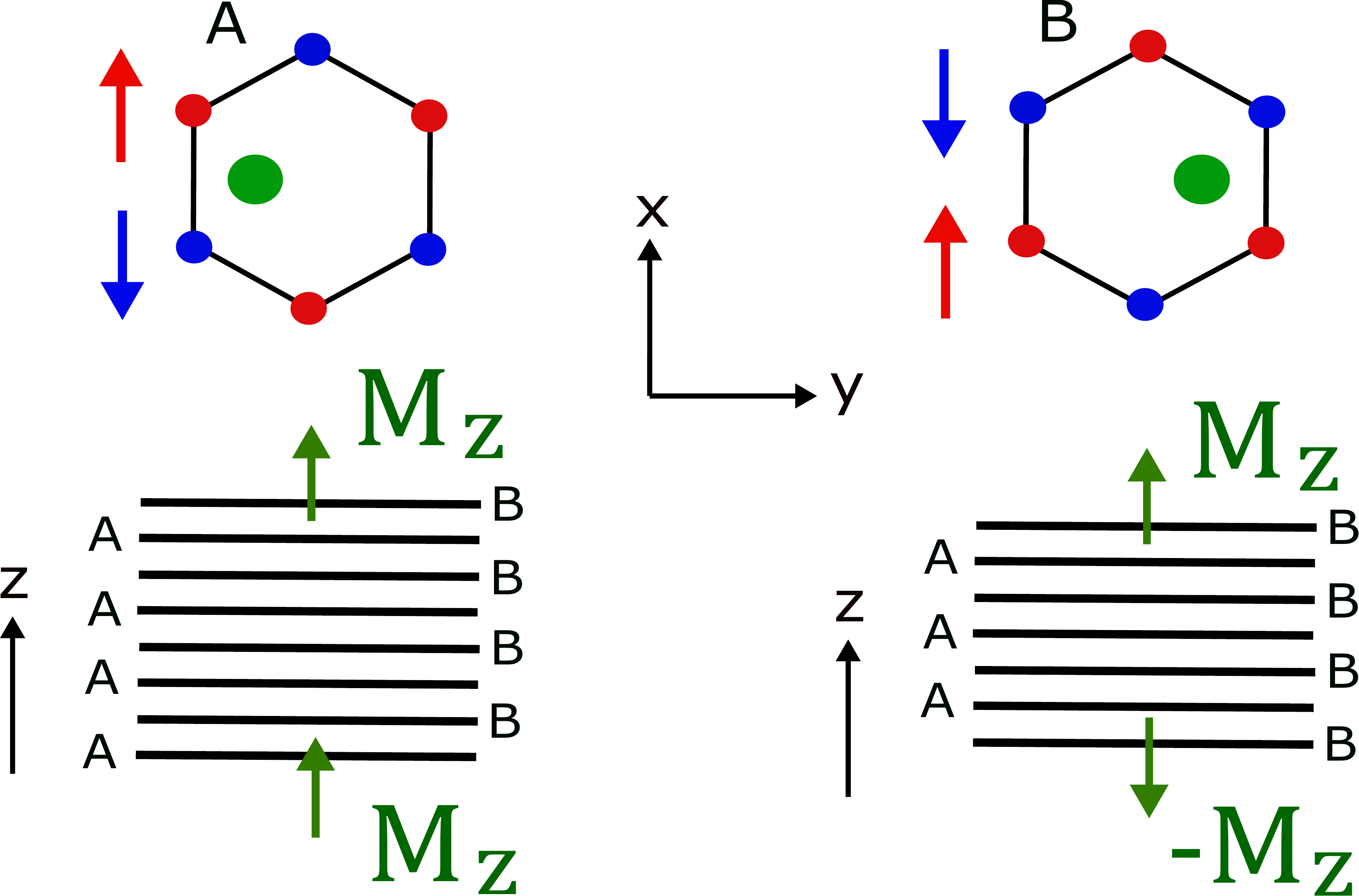} 

\protect\caption{Weak ferromagnet model. Two layers that form the unit cell of the system. 
The non-magnetic green atom is in the plane with red and blue sites. Each unit cell on its own has a $M_{z}L_{x}$ Dzyaloshinskii's invariant (see Figs. \ref{fig:fig2} and \ref{fig:fig3}).
A combination of mirror reflection in $xy$ and $xz$ planes and time-reversal allows for it.
However, in a bulk system, a combination of mirror reflection in the $xz$ plane and translation by half the unit cell in $z-$ direction prevents $M_{z}L_{x}$ from being the invariant in the bulk. This is because spin-orbit coupling needed for the magnetic moment to appear in each layer in the bulk is suppressed by adjacent top and bottom layers. Breaking this symmetry by truncating the crystal would allow for the surface weak ferromagnetism.}
\label{fig:fig1}  
\end{figure}
\begin{figure}[t] 
\includegraphics[width=0.8 \columnwidth ]{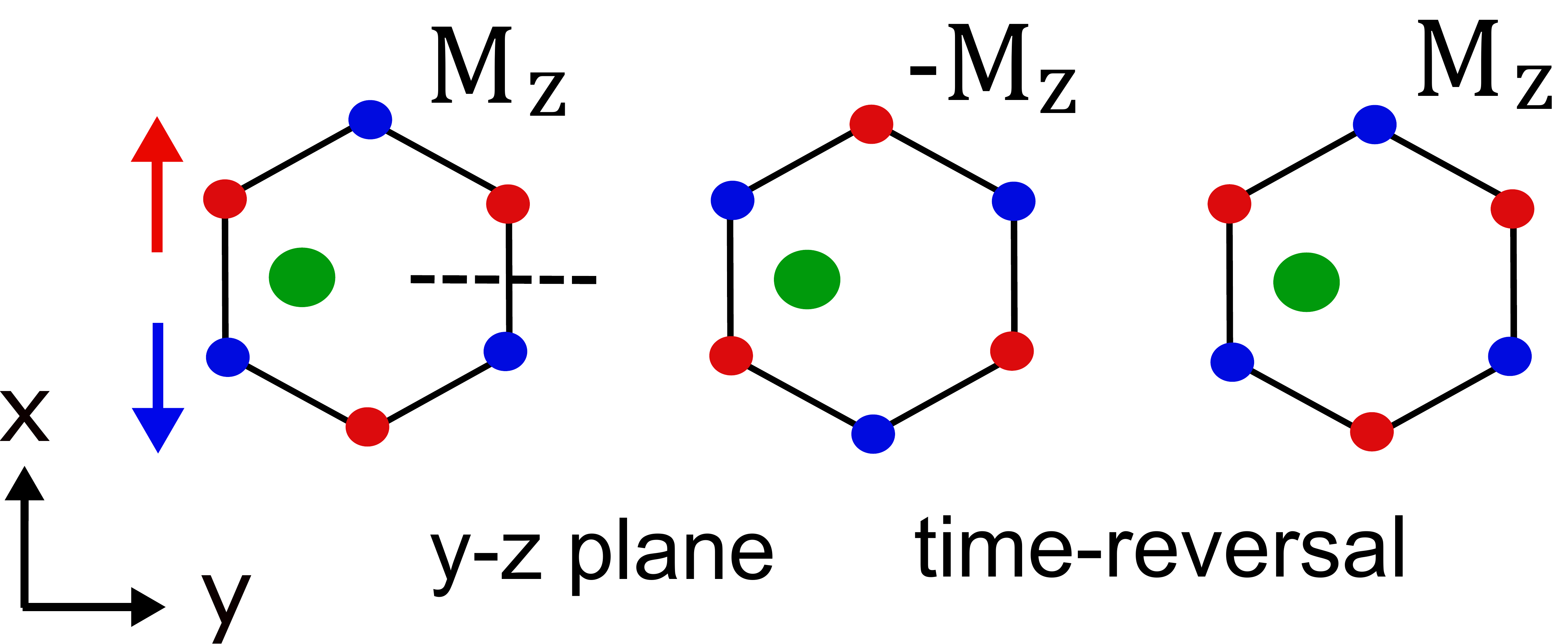} 

\protect\caption{A single layer of a weak ferromagnet is shown. Symmetry operation that connects the two magnetic sublattices consists of a 
mirror reflection in the $yz$ plane and time-reversal. The spins don't change under the mirror operation, however exchange of sites due to the reflection changes sign of $L_{x}$. The sign of $M_{z}$ is also changed under the mirror reflection. Time-reversal restores the sign of $L_{x}$ and $M_{z}$.
However, a mirror reflection in $xy$ plane flips the sign of $L_{x}$ while keeps $M_{z}$ intact. Therefore, for the $M_{z}L_{x}$ Dzyaloshinskii's invariant to occur, one has to break the mirror symmetry in $xy$ plane. This is exactly what happens at the surface of the layered antiferromagnet shown in Fig. (\ref{fig:fig1}).}
\label{fig:fig2}  
\end{figure}

\begin{figure}[h] 
\includegraphics[width=0.5 \columnwidth ]{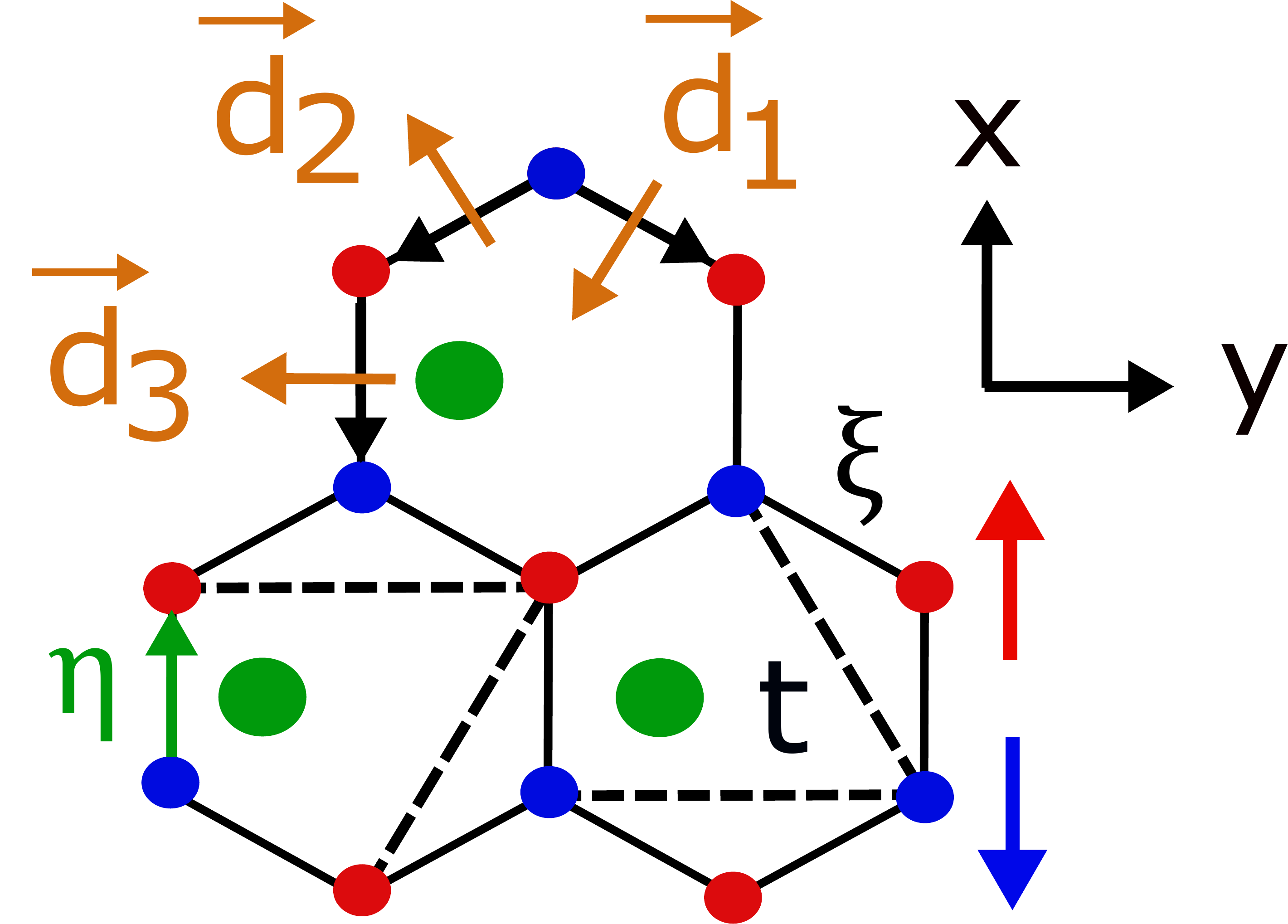} 

\protect\caption{Solid black lines represent nearest-neighbor hopping $\xi$, while dashed lines describe second-nearest neighbor anisotropic hopping $t$. It is the breaking of the $xy$ plane mirror reflection symmetry that turns mirror-symmetric genuine N\'{e}el ordered antiferromagnet into a weak ferromagnet.
Orange arrows are the directions of the spin-orbit coupling of the Rashba type due to the broken $xy$ plane mirror reflection symmetry, ${\bf d}_{1} = \frac{1}{2}(-\sqrt{3},-1)$, ${\bf d}_{2} = \frac{1}{2}(\sqrt{3},-1)$, ${\bf d}_{3} = (0,-1)$. Parameter $\eta$ is the out-of-plane spin-orbit coupling the green atom creates on the link. The sign of $\eta$ is chosen in accord with the green arrow that describes direction of hopping.
The N\'{e}el order depicted by red and blue circles and arrows is in $x$ direction.  }
\label{fig:fig3}  
\end{figure}

\begin{figure}[t] 
\includegraphics[width=0.95 \columnwidth ]{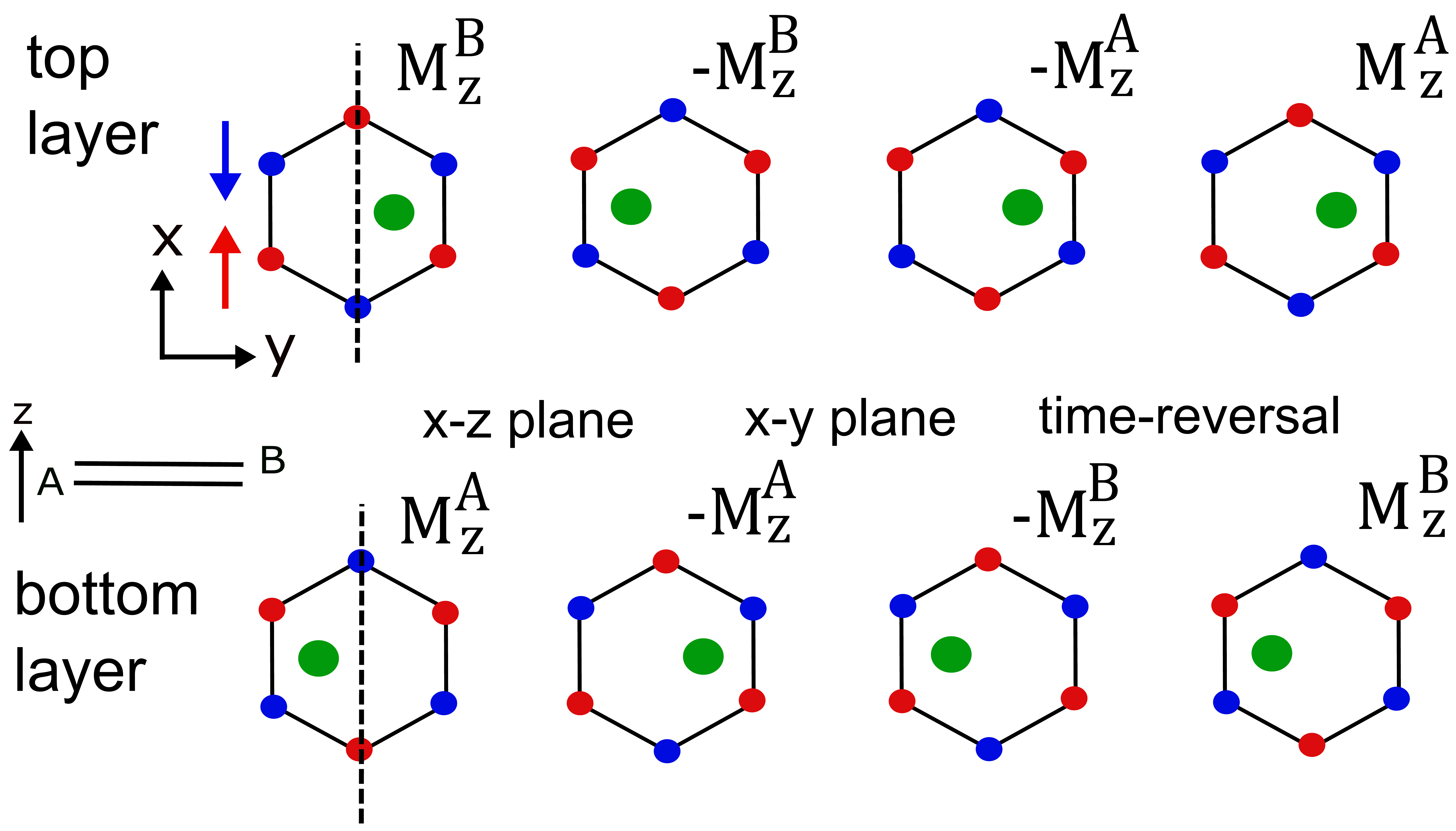} 

\protect\caption{Weak ferromagnet model. 
A symmetry operation that connects the two layers, top and bottom, of the unit cell is shown.
The operation requires $M_{z}^{\mathrm{A}} = M_{z}^{\mathrm{B}}$. This is also confirmed in the text by the Berry curvature calculation.}
\label{fig:fig4}  
\end{figure}

We introduce the system shown in Fig. (\ref{fig:fig1}). 
The unit cell of the system is composed of two layers labeled by A and B in the regular AA stacking way. 
Each layer has a N\'{e}el order described in Fig. (\ref{fig:fig1}) by red and blue sites.
The interlayer interaction is assumed to be antiferromagnetic, however, our arguments below will only be slightly changed if the interaction is ferromagnetic.
The unit cell is then translated in $z-$ direction to fill the whole space. The distance between neighbor layers is assumed to be the same.
The Hamiltonian of fermions in the proposed system is
\begin{align}\label{Hamiltonian}
\hat{H}_{{\bf k}} = \left[
\begin{array}{cc} \hat{H}_{\mathrm{A};{\bf k}} & \zeta_{\bf k}\hat{\sigma}_{0} \\
\zeta_{\bf k}^{*}\hat{\sigma}_{0} & \hat{H}_{\mathrm{B};{\bf k}}
\end{array}
\right],
\end{align}
where layer A Hamiltonian is
\begin{align}\label{LayerA}
\hat{H}_{\mathrm{A};{\bf k}}= 
\left[
\begin{array}{cc} 
{\bf m}\cdot{\bm \sigma} - t_{{\bf k}}  & \xi_{\bf k} + i {\bm \gamma}_{\bf k}\cdot {\bm \sigma}\\
\xi_{\bf k}^{*} - i {\bm \gamma}_{\bf k}^{*}\cdot {\bm \sigma} & -{\bf m}\cdot{\bm \sigma} + t_{{\bf k}}
\end{array}
\right],
\end{align}
and layer B Hamiltonian is
\begin{align}\label{LayerB}
\hat{H}_{\mathrm{B};{\bf k}}= 
\left[
\begin{array}{cc} 
-{\bf m}\cdot{\bm \sigma} + t_{{\bf k}}  & \xi_{\bf k} - i {\bm \gamma}_{\bf k}\cdot {\bm \sigma} \\
\xi_{\bf k}^{*} + i {\bm \gamma}_{\bf k}^{*}\cdot {\bm \sigma} & {\bf m}\cdot{\bm \sigma} - t_{{\bf k}}
\end{array}
\right],
\end{align}
where ${\bf m}\cdot {\bm \sigma}$ is the interaction between fermions and the N\'{e}el order,
$
t_{\bf k} = 2t\sin\left(\frac{\sqrt{3}k_{x}}{2}\right)\sin\left(\frac{k_{y}}{2}\right) $
is the anisotropic hopping within the sublattices,
$
\xi_{\bf k} =  \xi e^{i\frac{k_{x}}{2\sqrt{3}}} \left[2 \cos\left( \frac{k_{y}}{2} \right) +  e^{-i \frac{\sqrt{3}}{2}k_{x} } \right]
$
is the first-nearest neighbor hopping. 
The structure of the spin-orbit coupling of a unit cell composed of just two layers, in each of which there is a broken $xy$ plane mirror symmetry, is depicted in Fig. (\ref{fig:fig3}). 
Rashba spin-orbit coupling is given by three vectors ${\bf d}_{1}$, ${\bf d}_{2}$, and ${\bf d}_{3}$.
The spin-orbit coupling should be understood in the tight-binding description as $i\lambda \sum_{ij} ({\bf d}_{ij}\cdot{\bm \sigma})_{\alpha\beta}c_{\alpha;i}^{\dag}c_{\beta;j}$, where ${\bf d}_{ij} = {\bf d}_{1,2,3}$ for appropriate fermion hoppings defined by indeces $i$ and $j$.
Fourier tranform of the spin-orbit coupling reads
$
\gamma_{\bf k}^{x} = i \sqrt{3}\lambda e^{i\frac{k_{x}}{2\sqrt{3}}}\sin\left( \frac{k_{y}}{2} \right),
$
$
\gamma_{\bf k}^{y} = - \lambda\left[ 2 e^{i\frac{k_{x}}{2\sqrt{3}}}\cos\left( \frac{k_{y}}{2} \right) +   e^{-i\frac{k_{x}}{\sqrt{3}}}\right].
$
There is also a spin-orbit coupling $\eta$ due to the position of the green atom shown in Fig. (\ref{fig:fig3}), it is given by
$
\gamma_{\bf k}^{z} =   \eta e^{-i\frac{k_{x}}{\sqrt{3}}}.
$
Finally, hopping matrix element between the layers is
$
\zeta_{\bf k} = \zeta/2 
$. 
In the bulk, when the unit cell is translated along the $z-$ axis, tight-binding parameters become $\zeta_{\bf k} =\zeta\cos(k_{z})$ and $\gamma_{\bf k}^{x} = \gamma_{\bf k}^{y} = 0$. The spin-orbit coupling $\lambda$ in a given layer of the three dimensional stacked system gets cancelled due to the compensation of the electric fields from the top and bottom layers.
The spin-orbit coupling $\eta$ doesn't get affected in each layer because it is due to the green atom.


In order to understand the magnetization in the bulk and on the surface of the system, it is sufficient to understand the magnetic structure of a single layer. Single layer is shown in Fig. (\ref{fig:fig2}). If the green atom was in the center of the hexagon, then the system would have possesed a combination of inversion and time-reversal among the symmetries. The inversion flips the sign of the N\'{e}el order, but keeps intact any possible magnetization in the system. Further application of the time-reversal restores the N\'{e}el order, while flips the sign of the magnetization. Therefore, magnetization is absent in the N\'{e}el ordered system with inversion and time-reversal. In order to have non-zero magnetization we break the inversion by changing the position of the green atom as shown in Figs. (\ref{fig:fig1}) and Fig. (\ref{fig:fig2}). The symmetry that connects the two magnetic sublattices is shown in Fig. (\ref{fig:fig2}) and is a combination of a mirror reflection in the $yz$ plane, that crosses the link in the cented, and time-reversal. Such an antiferromagnet is called the mirror-symmetric antiferromagnet and was found in \cite{Zyuzin2025a}. The symmetry would have allowed for a magnetization pointing in $z-$direction. However a combination of a mirror reflection in $xy$ plane and time-reversal, that restores the system, eliminates any possible magnetization. Thus, this symmetry must be broken for magnetization to appear.


Let us now follow the steps of \cite{GolubinskiiZyuzin2026} and derive the Berry curvature of conducting fermions in the single layer shown in Fig. (\ref{fig:fig3}).
For example, let us consider $\hat{H}_{\mathrm{A};{\bf k}}$ with the $\hat{\Psi} = (\Psi_{\mathrm{R}\uparrow},\Psi_{\mathrm{R}\downarrow},\Psi_{\mathrm{B}\uparrow},\Psi_{\mathrm{B}\downarrow})^{\mathrm{T}}$ spinor structure, where R/B is the red/blue sites, and the arrows stand for spin. The N\'{e}el order ${\bf m}$ is defined by two angles, $\phi$ and $\theta$, i.e. ${\bf m} =m \left[\cos(\phi)\sin(\theta),\sin(\phi)\sin(\theta),\cos(\theta)\right]$.
As suggested in \cite{GolubinskiiZyuzin2026} it is convenient to transform the Hamilotnian to the conduction and valence bands by first rotating the spin axis such that N\'{e}el order is $\hat{T}^{-1} {\bf m}\cdot{\bm \sigma} \hat{T} = m\sigma_{z}$, where $m = \sqrt{m_{x}^2+m_{y}^2+m_{z}^2}$. Then we restucture the Hamiltonian to conduction and valence bands given by the $\hat{\tilde{\Psi}} = (\tilde{\Psi}_{\mathrm{R}\uparrow},\tilde{\Psi}_{\mathrm{B}\downarrow},\tilde{\Psi}_{\mathrm{R}\downarrow},\tilde{\Psi}_{\mathrm{B}\uparrow})^{\mathrm{T}}$ space,
\begin{align}
\hat{H}_{\mathrm{A};{\bf k}} = \left[ \begin{array}{cc} \hat{H}^{\mathrm{c}} & \hat{C} \\ \hat{C}^{\dag} & \hat{H}^{\mathrm{v}} \end{array}\right],
~~
 \hat{C} =  \left[ \begin{array}{cc} 0 & \xi_{\bf k} + ia_{\bf k}  \\  \xi_{\bf k}^{*} + ia^{*}_{\bf k} & 0 \end{array}\right],
\end{align}
where Hamiltonians describing conduction (c) and valence (v) bands are
\begin{align}\label{conductionWeak}
\hat{H}^{\mathrm{c}} = \left[ \begin{array}{cc} m-t_{\bf k} & ib_{{\bf k}} \\
-ib_{{\bf k}}^{*}  & m+t_{\bf k} \end{array} \right],  \hat{H}^{\mathrm{v}} = \left[ \begin{array}{cc} -m - t_{\bf k} & i\bar{b}_{{\bf k}}  \\
-i\bar{b}_{{\bf k}}^{*} & -m + t_{\bf k} \end{array} \right],
\end{align}
where we have introduced
\begin{align}\label{ak12ak}
&
a_{\bf k} = \gamma^{z}_{\bf k} \cos(\theta) + \frac{1}{2} \sin\left( \theta\right)  \left(\gamma^{-}_{\bf k}e^{i\phi}+ \gamma^{+}_{\bf k}e^{-i\phi}\right) ,
\\
&
b_{{\bf k}} = -\gamma^{z}_{\bf k} \sin(\theta) +\gamma^{-}_{\bf k} \cos^2\left( \frac{\theta}{2}\right) e^{i\phi} - \gamma^{+}_{\bf k} \sin^2\left( \frac{\theta}{2}\right)e^{-i\phi}  ,
\nonumber
\\
&
\bar{b}_{{\bf k}} = -\gamma^{z}_{\bf k} \sin(\theta) +\gamma^{+}_{\bf k} \cos^2\left( \frac{\theta}{2}\right) e^{-i\phi} - \gamma^{-}_{\bf k} \sin^2\left( \frac{\theta}{2}\right)e^{i\phi},
\nonumber
\end{align}
where $\gamma_{\bf k}^{\pm} = \gamma_{\bf k}^{x} \pm i \gamma_{\bf k}^{y}$, where note that $\gamma_{\bf k}^{x/y}$ are in general complex, hence the need for $b_{{\bf k}}$ and $\bar{b}_{{\bf k}}$. For the N\'{e}el order in $x-$direction, $\phi = 0$ and $\theta = \pi/2$, the parameters are $b_{{\bf k}} = - \gamma_{\bf k}^{z} - i \gamma_{\bf k}^{y}$, and $a_{\bf k} = \gamma_{\bf k}^{x}$. We set the chemical potential to the conduction band and study the Berry phase of fermions described by $\hat{H}^{\mathrm{c}}$. The part of the Berry curvature that doesn't vanish after the integration over the Brilluoin zone is  
\begin{align}\label{Berry}
\Omega^{\pm}_{xy;{\bf k}}
 =
 \pm \frac{\lambda \eta t m_{x}}{8\sqrt{3} m}\frac{1}{\left( t_{\bf k}^2 + \vert b_{\bf k}\vert^2 \right)^{3/2}} ,
\end{align}
where $m_{x}/m$ was restored to highlight the dependence of the Berry curvature on the $m_{x}$.
The $\pm$ corresponds to the two subbands of the conduction band. The Berry curvature is the manifistation of the orbital magnetization of conduction fermions. By examening the structure of the Berry curvature Eq. (\ref{Berry}), we observe that the magnetization depends on the sign of a product of four parameters. These are $m_{x}$, $t$, $\eta$, and $\lambda$. We can then, for example, relate the magnetization of a layer A to the magnetization of a layer B when the two are placed together to form a unit cell, see Fig. (\ref{fig:fig1}) for details.


Now we analyze the unit cell shown in Fig. (\ref{fig:fig4}) using symmetry arguments. Layer B can be obtained from layer A by applying a mirror reflection across the \(xz\) plane that passes through the sites of the hexagon. This reflection flips the spins pointing along the \(x\)-direction and interchanges the positions of the green atoms. If layer A possesses a magnetization \(\mathbf{M}_{\mathrm{A}}\) pointing along the \(z\)-direction (i.e., in the \(+M_{\mathrm{A};z}{\bf e}_z\) direction), this mirror operation reverses its sign, causing it to point along \(-M_{\mathrm{A};z}{\bf e}_{z}\).Next, to exchange the layer positions within the unit cell, one must apply a mirror reflection across the \(xy\) plane. This operation flips the spins of the N\'{e}el order within each layer but leaves the magnetization \(\mathbf{M}_{\mathrm{A}}\) intact. Consequently, to map the unit cell back onto itself, we must finally apply the time-reversal symmetry operation. Time reversal reverses the signs of all spins and the magnetization, restoring \(-M_{\mathrm{A};z}{\bf e}_z\) to \(+M_{\mathrm{A};z}{\bf e}_z\). Thus, we find that the magnetization of layer B is identical in both magnitude and direction to that of layer A. This conclusion can be independently verified by examining the Berry curvature in Eq.~(\ref{Berry}). The Hamiltonian of layer B is obtained from that of layer A, given in Eq.~(\ref{LayerA}), by reversing the signs of \(m_{x}\), \(t\), \(\eta \), and \(\lambda \). As a result, the Berry curvature of the fermions in layer B shares the same sign as in layer A, confirming that the magnetizations of both layers are indeed equivalent to each other.

With a clear understanding of the system's unit cell, we can now readily explain the emergence of surface weak ferromagnetism in our model. In the bulk, the unit cell is translated along the \(z\)-direction. Because the spin-orbit coupling \(\lambda \) from individual layers effectively cancels out within the bulk, net bulk magnetization vanishes. More formally, a combined symmetry operation—consisting of a translation by half the unit cell distance along the \(z\)-axis and a mirror reflection across the \(xz\) plane (without time-reversal)—explicitly forbids magnetization in the bulk system.However, structural boundaries such as surfaces or dislocations break this protective symmetry, thereby lifting the restriction and permitting a weak ferromagnetic moment to develop along the \(z\)-direction. This physical scenario is illustrated in Fig.~\ref{fig:fig1} for various surface terminations.

It is important to note that it is not that there is a $M_{z} \neq 0$ in each layer, and the system is composed of layers with the altering sign of $M_{z}$ so as to have a total of $M_{z} = 0$ in the bulk. For example, such models were introduced in  \cite{Yakovenko2016,Zyuzin2024} and many more. Or, for example, this might be the case with MnTe antiferromagnet, where each layer is ferromagnetic and the N\'{e}el order is between the layers. 
 Contrary to that, in our model each layer on its own has $M_{z} = 0$.
This is because in-plane mirror symmetry isn't broken in a single layer. When the two layers are put together to form a unit cell, the mirror symmetry gets broken, and as a result each layer acquires $M_{z}\neq 0$ of the same sign.
The same happens when a single layer is put on the substrate that creates an uncompensated electric field in the layer.
Alternatively a single layer can be put between gates such that one can control the magnetization direction in the layer by switching polarity of the gates.
Technicaly, the broken mirror symmetry creates a non-zero spin-orbit coupling $\lambda \neq 0$ that is responsible for the appearence of the magnetic moment. This is essentially the mechanism of surface weak ferromagnetism.

Let us try to identify interesting effects associated with the surface weak ferromagnetism. 
First of all there is going to be anomalous Hall effect occurring only on the surfaces of the antiferromagnet. 
Furthermore, one can come up with a model when opposite surfaces have the magnetic moments pointing in opposite directions. 
As a result, equal sign polar Kerr effect upon reflection from the opposite surfaces and absence of the Faraday effect is expected.
Such a situation has been experimentally observed in the pseudogap of cuprates.
Moreover, we can think of a model of antiferromagnetic superconductor that may have unpaired metallic surfaces due to the surface weak ferromagnetism. Spin-singlet s-wave superconducting pairing is expected to be between the layers due to the spin-splitting in each layer. As a result, the superconductivity can be suppressed near the surfaces due to the surface magnetic moment and there is a possibility of having an unpaired fermion liquid on the surfaces.

The author is grateful to Pirinem School of Theoretical Physics for hospitality. 
This work is supported by FFWR-2024-0016.

\end{document}